\documentclass[3p,times,procedia,longtitle]{elsarticle}
\usepackage{ecrc}

\volume{00}

\firstpage{1}

\journalname{Physics Procedia}

\runauth{}

\jid{procs}

\jnltitlelogo{Physics Procedia}

\CopyrightLine{2014}{Published by Elsevier Ltd.}

\usepackage{amssymb}
\usepackage[figuresright]{rotating}

\begin{document}

% Definitions...
\def\nuc#1#2{${}^{#1}$#2}
\def\BBz{0$\nu\beta\beta$}
\def\BBt{2$\nu\beta\beta$}
\def\BB{$\beta\beta$}
\def\Tz{$T^{0\nu}_{1/2}$}
\def\Tt{$T^{2\nu}_{1/2}$}
\def\mj{M{\sc ajo\-ra\-na}}
\def\dem{D{\sc e\-mon\-strat\-or}}
\def\mg{M{\sc a}G{\sc e}}
\def\geant{G{\sc eant}4}
\def\QBB{Q$_{\beta\beta}$}
\def\mBB{$\left < \mbox{m}_{\beta\beta} \right >$}
\def\ge{$^{76}$Ge}

\begin{frontmatter}



\dochead{}
%% e.g. \dochead{17th International Conference on Dynamical Processes in Excited States of Solids}

\title{Testing the Ge detectors for the  \mj\ \dem}

\author[lanl]{W.~Xu}  
\author[lbnl]{N.~Abgrall}		
\author[pnnl]{E.~Aguayo}
\author[usc,ornl]{F.T.~Avignone~III}
\author[ITEP]{A.S.~Barabash}	
\author[ornl]{F.E.~Bertrand}
\author[lanl]{M.~Boswell} 
\author[JINR]{V.~Brudanin}
\author[duke,tunl]{M.~Busch}	
\author[usd]{D.~Byram} 
\author[sdsmt]{A.S.~Caldwell}
\author[lbnl]{Y-D.~Chan}
\author[sdsmt]{C.D.~Christofferson} 
\author[ncsu,tunl]{D.C.~Combs}  
\author[uw]{C. Cuesta}	
\author[uw]{J.A.~Detwiler}	
\author[uw]{P.J.~Doe} 
\author[ut]{Yu.~Efremenko}
\author[JINR]{V.~Egorov}
\author[ou]{H.~Ejiri}
\author[lanl]{S.R.~Elliott}
\author[pnnl]{J.E.~Fast}
\author[unc,tunl]{P.~Finnerty}  
\author[unc,tunl]{F.M.~Fraenkle} 
\author[ornl]{A.~Galindo-Uribarri}	
\author[unc,tunl]{G.K.~Giovanetti}  
\author[lanl]{J. Goett}	
\author[ornl]{M.P.~Green}  
\author[uw]{J. Gruszko}		
\author[usc]{V.E.~Guiseppe}	
\author[JINR]{K.~Gusev}
\author[alberta]{A.L.~Hallin}
\author[ou]{R.~Hazama}
\author[lbnl]{A.~Hegai\fnref{TU}} 
\author[unc,tunl]{R.~Henning}
\author[pnnl]{E.W.~Hoppe}
\author[sdsmt]{S. Howard}  
\author[unc,tunl]{M.A.~Howe}
\author[blhill]{K.J.~Keeter}
\author[ttu]{M.F.~Kidd}	
\author[JINR]{O.~Kochetov}
\author[ITEP]{S.I.~Konovalov}
\author[pnnl]{R.T.~Kouzes}
\author[pnnl]{B.D.~LaFerriere}   
\author[uw]{J.~Leon}	
\author[ncsu,tunl]{L.E.~Leviner}
\author[sjtu]{J.C.~Loach}	
\author[unc,tunl]{J.~MacMullin}
\author[unc,tunl]{S.~MacMullin} 
\author[usd]{R.D.~Martin}
\author[unc,tunl]{S. Meijer}	
\author[lbnl]{S.~Mertens}		
\author[ou]{M.~Nomachi}
\author[pnnl]{J.L.~Orrell}
\author[unc,tunl]{C. O'Shaughnessy}	
\author[pnnl]{N.R.~Overman}  
\author[ncsu,tunl]{D.G.~Phillips~II}  
\author[lbnl]{A.W.P.~Poon}
\author[usd]{K.~Pushkin} 
\author[ornl]{D.C.~Radford}
\author[unc,tunl]{J.~Rager}	
\author[lanl]{K.~Rielage}
\author[uw]{R.G.H.~Robertson}
\author[ut,ornl]{E.~Romero-Romero} 
\author[lanl]{M.C.~Ronquest}	
\author[uw]{A.G.~Schubert}		
\author[unc,tunl]{B.~Shanks}	
\author[ou]{T.~Shima}
\author[JINR]{M.~Shirchenko}
\author[unc,tunl]{K.J.~Snavely}	
\author[usd]{N.~Snyder}	
\author[sdsmt]{A.M.~Suriano} 
\author[blhill,sdsmt]{J.~Thompson} 
\author[JINR]{V.~Timkin}
\author[duke,tunl]{W.~Tornow}
\author[unc,tunl]{J.E.~Trimble}
\author[ornl]{R.L.~Varner}  
\author[ut]{S.~Vasilyev}
\author[lbnl]{K.~Vetter\fnref{ucb}}
\author[unc,tunl]{K.~Vorren} 
\author[ornl]{B.R.~White}	
\author[unc,tunl,ornl]{J.F.~Wilkerson}    
\author[usc]{C.~Wiseman}		
\author[JINR]{E.~Yakushev}
\author[ncsu,tunl]{A.R.~Young}
\author[ornl]{C.-H.~Yu}
\author[ITEP]{V.~Yumatov}

\address[lanl]{Los Alamos National Laboratory, Los Alamos, NM, USA}
\address[lbnl]{Nuclear Science Division, Lawrence Berkeley National Laboratory, Berkeley, CA, USA}
\address[pnnl]{Pacific Northwest National Laboratory, Richland, WA, USA}
\address[usc]{Department of Physics and Astronomy, University of South Carolina, Columbia, SC, USA}
\address[ornl]{Oak Ridge National Laboratory, Oak Ridge, TN, USA}
\address[ITEP]{Institute for Theoretical and Experimental Physics, Moscow, Russia}
\address[JINR]{Joint Institute for Nuclear Research, Dubna, Russia}
\address[duke]{Department of Physics, Duke University, Durham, NC, USA}
\address[tunl]{Triangle Universities Nuclear Laboratory, Durham, NC, USA}
\address[usd]{Department of Physics, University of South Dakota, Vermillion, SD, USA} 
\address[sdsmt]{South Dakota School of Mines and Technology, Rapid City, SD, USA}
\address[ncsu]{Department of Physics, North Carolina State University, Raleigh, NC, USA}
\address[uw]{Center for Experimental Nuclear Physics and Astrophysics, and Department of Physics, University of Washington, Seattle, WA, USA}
\address[ut]{Department of Physics and Astronomy, University of Tennessee, Knoxville, TN, USA}
\address[ou]{Research Center for Nuclear Physics and Department of Physics, Osaka University, Ibaraki, Osaka, Japan}
\address[unc]{Department of Physics and Astronomy, University of North Carolina, Chapel Hill, NC, USA}
\address[alberta]{Centre for Particle Physics, University of Alberta, Edmonton, AB, Canada}
\address[blhill]{Department of Physics, Black Hills State University, Spearfish, SD, USA} 
\address[ttu]{Tennessee Tech University, Cookeville, TN, USA}
\address[sjtu]{Shanghai Jiao Tong University, Shanghai, China}
\fntext[TU]{Permanent address: Tuebingen University, Tuebingen, Germany}
\fntext[ucb]{Alternate Address: Department of Nuclear Engineering, University of California, Berkeley, CA, USA}

\begin{abstract}
High purity germanium (HPGe) crystals will be used for the \mj\ \dem, where they serve as both the source and the detector for neutrinoless double beta decay. It is crucial for the experiment to understand the performance of the HPGe crystals. A variety of crystal properties are being investigated, including basic properties such as energy resolution, efficiency, uniformity, capacitance, leakage current and crystal axis orientation, as well as more sophisticated properties, \textit{e.g.} pulse shapes and dead layer and transition layer distributions. In this paper, we will present our measurements that characterize the HPGe crystals. We will also discuss our simulation package for the detector characterization setup, and show that additional information can be extracted from data-simulation comparisons.
\end{abstract}


\end{frontmatter}

% \linenumbers

\section{Introduction to neutrinoless double beta decay}
Two-neutrino double beta decay (\BBt) is a rare process where two neutrons simultaneously decay into protons and emit two electrons and two antineutrinos. It only takes place when single beta decay is energetically forbidden or greatly inhibited. After its first direct detection in 1987, this process has been observed in several isotopes with even numbers of protons and neutrons~\cite{BBt1,BBt2}. If neutrinos are Majorana particles, {\it i.e.} particles that are their own anti-particles, then the neutrino could be exchanged as a virtual particle between two neutrons, and therefore the so called neutrinoless double beta decay (\BBz) could occur. The observation of \BBz\ would directly prove that the lepton number is violated, which would have profound implications on the mass asymmetry in the universe, as the lepton number is directly linked to the baryon number in many theories beyond the Standard Model. In addition to being the only practical way to establish Majorana neutrinos, \BBz\ experiments could also shed light on the absolute neutrino mass scale and neutrino mass hierarchy~\cite{BBz1, BBz2, BBz3}.
 
\section{The \mj\ \dem}
The goal of the \mj\ \dem~\cite{MJD1}, currently under construction 4850 feet underground in the Sanford Underground Research Facility (SURF), is to demonstrate the techniques required for a definitive next-generation tonne scale
\BBz\ experiment with enriched Ge detectors, including achieving a low enough background and establishing the feasibility to construct and deploy modular arrays of Ge detectors. The \dem\ will also test the Klapdor-Kleingrothaus claim~\cite{KK1} of direct observation of \BBz\ decay and search for new physics beyond the Standard Model, such as low-mass dark matter and axions.\\

\begin{figure}
    \centering
    \includegraphics[width=0.7 \textwidth]{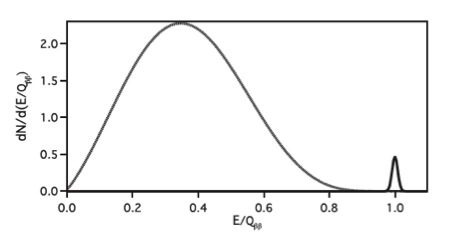}
    \caption{A sketch of the spectrum of the summed energy of the two electrons in double beta decays. The peak at E/Q=1 is from \BBz\ decays and the continuum to the left is from the intrinsic background of \BBt\ decays. The curves were drawn assuming that the decay rate of \BBt\ is 100 times higher than \BBz\ and the $1\sigma$ energy resolution is 2\%. Figure adapted from~\cite{BBz1}.}
    \label{fig:f1}
\end{figure}

A total of 40 kg of High-Purity Germanium (HPGe) detectors will be deployed in the \dem, composed of 30 kg of Ge crystals enriched to 86\% of \ge \ and 10 kg of natural Ge crystals. These crystals will serve as both the source and the detector for neutrinoless double beta decay. If neutrinos are Majoarana particles, the \ge\ nuclei could undergo \BBz\ decay and the summed energy of the two electrons emitted in the process would be at the Q-value of the decay, 2039~keV, providing the signature for the process. However, the \ge\ nuclei also undergo \BBt\ decay at a much higher rate, which introduces a continuum background in the spectrum of the summed electron energies up to very close to the Q-value. The excellent energy resolution of Ge detectors, about 0.2\% at the Q-value, reduces the Region of Interest (ROI) to merely 4~keV around the Q-value, effectively eliminating the contamination from the tail of the \BBt \ continuum. Figure~\ref{fig:f1} illustrates this effect.\\

\begin{figure} 
    \centering
    \includegraphics[width=0.9 \textwidth]{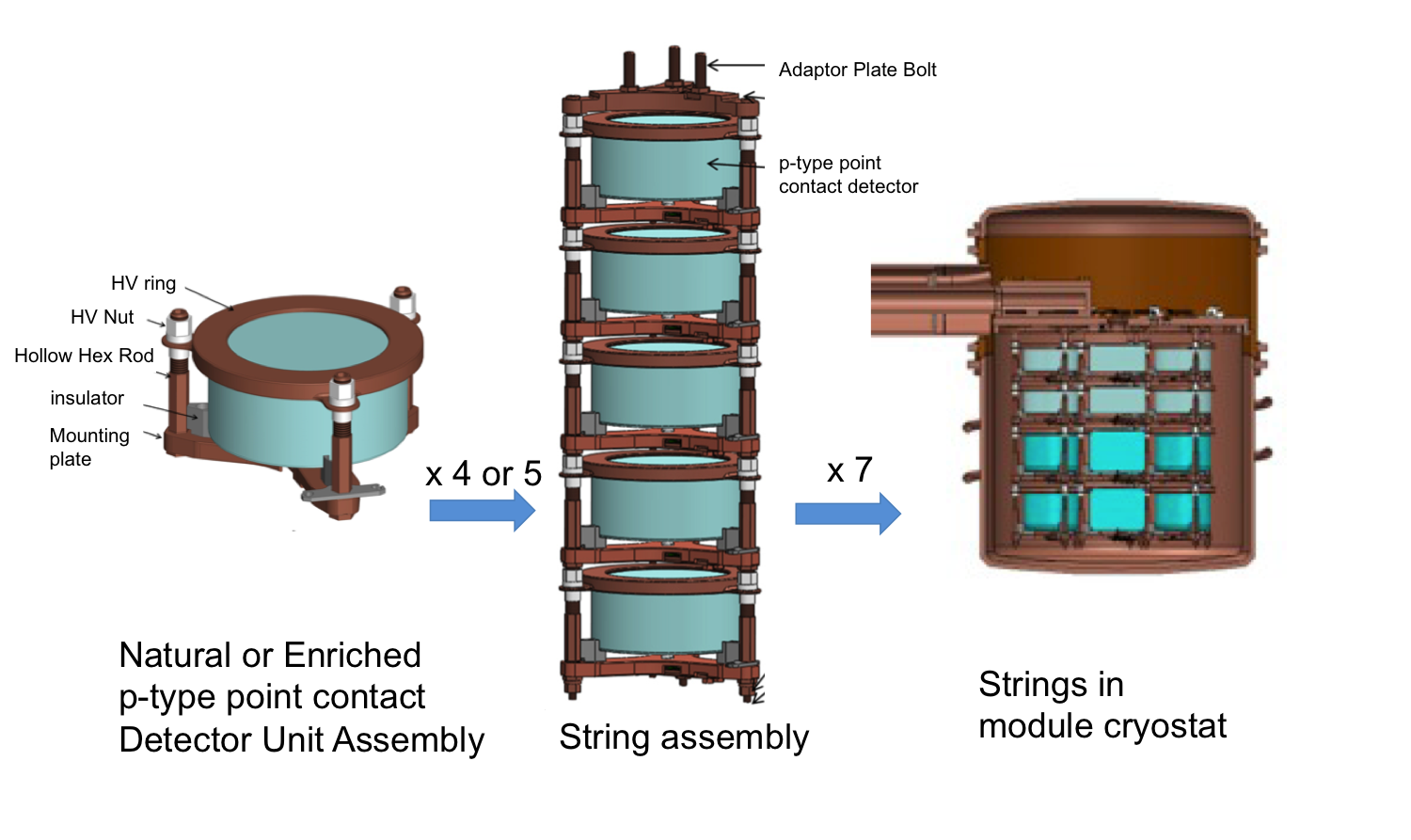}
    \caption{The modular approach of  \mj\ \dem. From left to right, the plot shows a detector unit assembly, a string assembly and finally the assembly of seven strings in a module cryostat.}
    \label{fig:g2}
\end{figure}

The \mj\ Collaboration choose to use a modular approach to construct the experimental apparatus. Four or five individual HPGe detectors and their associated low-mass low-radioactivity mounting structures and electronics are stacked together into one string assembly, and seven string assemblies are installed into one cryostat, as shown in Figure~\ref{fig:g2}. The \mj\ \dem\ will have two independent cryostats with fourteen strings of detectors. Such an approach has several advantages, for example, data taking can start immediately after the first cryostat is constructed, and more importantly, it is easier to scale up to a future large scale apparatus. The manipulation of bare Ge crystals as well as the construction of detector and string assemblies are always performed in custom built glove boxes, which are constantly purged by dry nitrogen air to protect the Ge crystals and minimize the presence of radon that could plate onto the detectors.

\section{P-type Point Contact Detectors}
\begin{figure} 
    \centering
    \includegraphics[width=0.9 \textwidth]{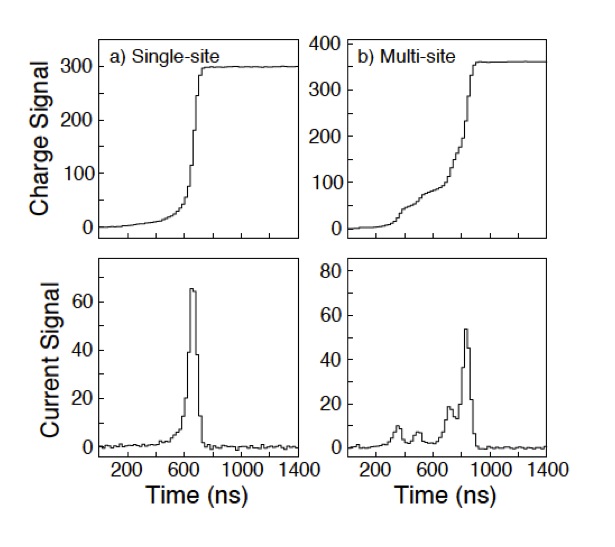}
    \caption{Current and charge pulse responses of a p-type point contact detector to single-site (a) and multi-site (b) events. The MS events exhibit multiple steps in the charge signal waveform and, correspondingly, multiple peaks in the current signal waveform. Figure adapted from~\cite{MJD1}. }
    \label{fig:g3}
\end{figure}

The choice of detector technology by \mj\ \dem\ is the p-type point contact (P-PC) HPGe detector~\cite{PPC1, PPC2}, which is optimal for the experiment largely thanks to its excellent pulse shape discrimination (PSD) ability. Comparing to the traditional coaxial Ge detectors, the spatially localized weighting potential inside the P-PC detectors effectively translates spatial separation of energy depositions into temporal separation. As a result, the waveforms for background events such as multiple Compton scatters exhibit structure corresponding to multiple energy deposition sites. Such events are called multiple-site (MS) events, and background $\gamma$ events are often MS events. In contrast, since electrons can only travel inside the Ge crystal for very short distances, no more than one millimeter, \BBz\ and \BBt\ decays are single-site (SS) events and their waveforms are characteristically different from the MS events, as shown in Figure~\ref{fig:g3}. By quantitatively analyzing the waveforms, the MS background can be very effectively eliminated. The P-PC detector also has small capacitance, which results in excellent energy resolution and low-energy threshold.

\section{MJD detector Acceptance and Characterization plan}
\begin{figure} 
    \centering
    \includegraphics[width=\textwidth]{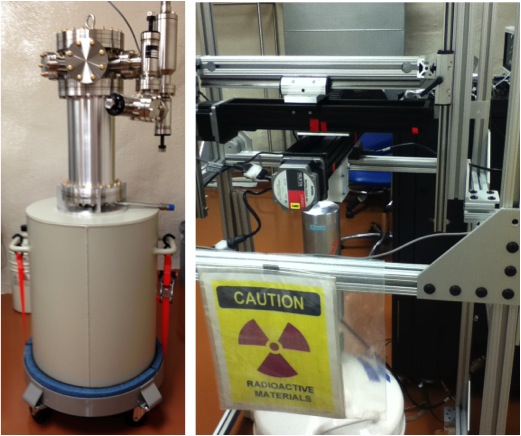}
    \caption{Photos of the setups used in detector and string tests. Left photo: A string test cryostat attached to a dewar that provides liquid nitrogen cooling. Right photo: A HPGe detector in its PopTop\texttrademark~cryostats is being calibrated using a collimated $^{133}$Ba source. The source is mounted on an automated x-y scanner system that allows scans across the top of the detector.  The scanner system can be reconfigured to allow scans in the z-direction.}
    \label{fig:setup}
\end{figure}

As the safeguard for the quality of Ge detectors received from the manufacture and as the initial phase of evaluating the detector performance, acceptance and characterization tests are conducted both at the vendor and in the SURF underground lab. We measure a suite of detector properties, including energy resolution, efficiency, leakage current, dead layer, peak shape and energy calibration linearity. The first three properties are the main criteria to accept the detectors. At the next step, string characterization tests are performed when several detectors are assembled into a string.  Custom built cryostats (string test cryostats) that can be introduced into the glove boxes are used for string storage and cooling during these tests, so that the strings always remain in dry nitrogen air or in vacuum. A photo of a string test cryostat is shown in the left panel of Figure~\ref{fig:setup}. Low-noise electronics, designed and manufactured by the collaboration, are used to extract the signals. The same string configurations and associated electronics will be used for the search of \BBz\ decay, so it is important to investigate the system performance at this stage with this setup. Finally, once the strings are installed into the \dem\ module cryostats, periodic calibration with various sources will be carried out through the data-taking period.\\

Most of the natural HPGe detectors for the \mj\ \dem\ are manufactured by Canberra~\cite{Canberra}, and they have already been delivered and tested previously. All of the enriched detectors and two natural detectors are manufactured by AMETEK/ORTEC~\cite{ORTEC}, and the production and delivery is still ongoing. The following sections will provide details of the acceptance and characterization tests for the ORTEC detectors at SURF (the tests for the Canberra detectors are similar). \\

After a group of HPGe detectors pass the tests at the vendor, they are transported to SURF in their individual PopTop\texttrademark~cryostats along with ORTEC preamps and HV filters, which are used in the initial tests. The Data Acquisition (DAQ) system consists of a NIM crate with a shaping amplifier and a Multiple Channel Analyzer (MCA), and a VME crate with a GRETINA Digitization card~\cite{GRE} and a HV module. The ORTEC preamps provide two identical signal outputs, one is fed into the Digitizer and the other is fed into the MCA after going through the shaping amplifier. A HV shutdown interface device from ORTEC is deployed as an interlock to automatically shut down the HV if the detector temperature rises.  A programmable pulser is used to study the response of the ORTEC preamps. An oscilloscope is used for various tasks, mainly monitoring the waveforms and measuring the baselines. The change of baseline at different bias voltages is then used to extract the leakage current, which is typically below 20 pA. The DAQ system for this test setup, and for \mj\ \dem, is controlled by the Object-Oriented Real-time Control and Acquisition (ORCA) software~\cite{ORCA}.\\

\begin{figure} 
    \centering
    \includegraphics[width=0.8 \textwidth]{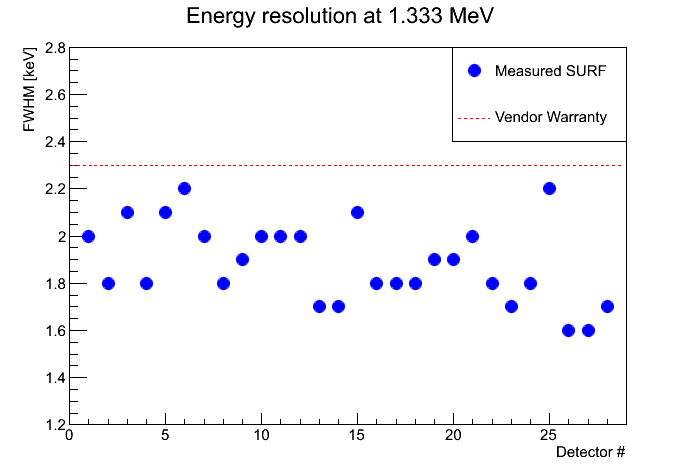}
    \caption{The energy resolution (FWHM) at 1332.5~keV of 28 HPGe detectors measured at SURF underground lab rooms. The uncertainty, not shown in the plot, is less than 0.1~keV. }
    \label{fig:FWHM}
\end{figure}

Flood measurements are performed with various sources placed 25 cm above the top surface of the PopTop\texttrademark~ cryostat. Data taken with $^{60}$Co sources are used to determine the detector energy resolution at 1332.5 keV and the detection efficiency relative to a $3\ in.\times 3\ in.$ NaI (Tl) detector. To sufficiently describe the detector responses, a Gaussian plus an exponential tail is used to fit the photo-peak and a step function plus a second order polynomial is used to describe the background around the photo-peak. The Full Width at Half Maximum (FWHM) of fitted Gaussian plus the tail is taken as the energy resolution. As of March 2014, we have received, tested and accepted 26 enriched detectors and 2 natural detectors from ORTEC. The measured energy resolutions at 1332.5~keV span from 1.6~keV to 2.2~keV, shown in Figure~\ref{fig:FWHM}, all better than our acceptance criterion. The onsite measurements of energy resolution and relative efficiency are consistent with the measurements conducted by the vendor. \\

Flood measurements taken with $^{133}$Ba and $^{214}$Am sources are used to extract the detector energy resolution at lower energies of 59.5 keV, 81.0 keV and 356.0 keV, so that the energy resolution as a function of the energy can be obtained. The $^{133}$Ba data are also used to extract the detector dead layer. Firstly the the ratio of number of events in the 81.0~keV photo-peak and in the 356.0~keV photo-peak is measured in the data. Then detailed \geant\ simulations of the PopTop\texttrademark~ cryostat and the test stand are carried out for each individual detector, with varying dead layer thickness. The same ratio in the simulation is calculated and compared to that in data, and the dead layer is obtained when these two ratios match. The nominal dead layer is about 1~mm and our extraction is usually consistent with the vendor specification within about 0.1 to 0.2~mm. The simulation geometry is part of  \mg\cite{MaGe}, which is a simulation framework based on \geant\, jointly developed by the \mj\ and the GERDA collaborations. In order to meet the specific physics and software requirements of the experiments, the \mg\ framework includes non-\geant\ physics processes such as generators of electronic waveforms in the HPGe, generic surface sampler, \BBz\ decays and more~\cite{MaGe}.\\

\begin{figure} 
    \centering
    \includegraphics[width=\textwidth]{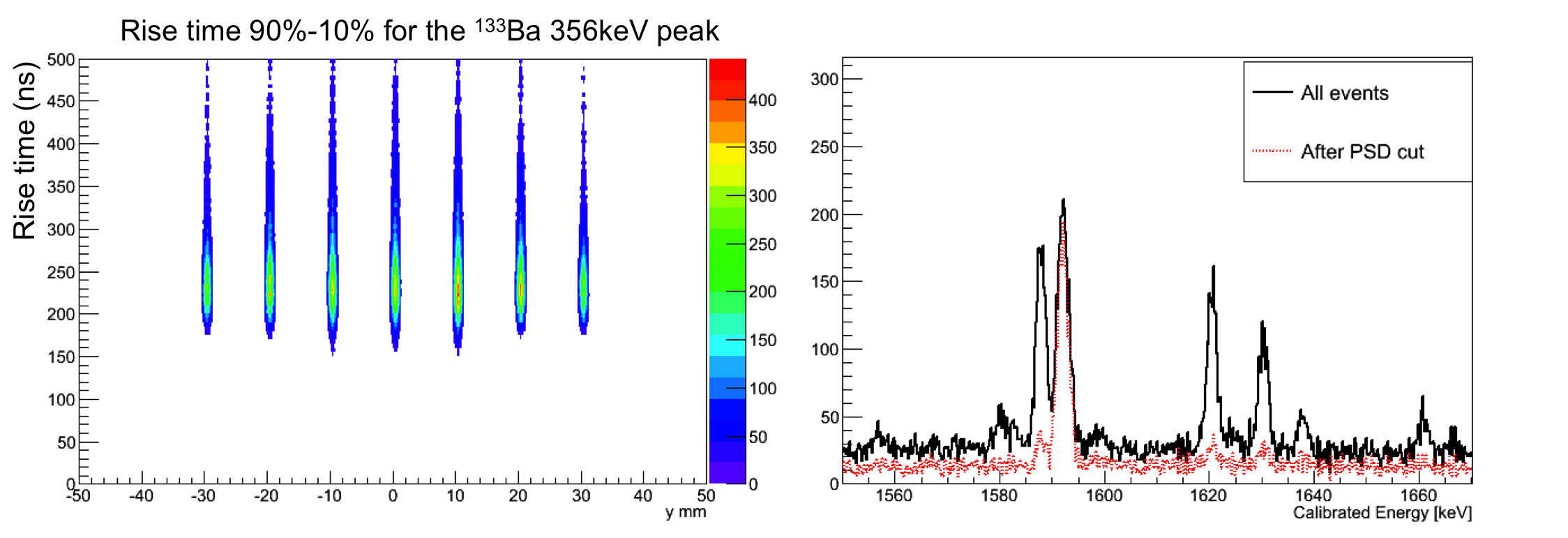}
    \caption{Left: The rise time of events around 356~keV, calculated as the time difference between 90\% of the signal and 10\% of the signal, is measured for each scan location across the top surface of a detector. Right: A typical spectrum around 1600 keV of a $^{232}$Th source, before (black solid line) and after (red dotted line) applying the A/E cut. The full energy peaks are largely reduced by the A/E cut.}
    \label{fig:results}
\end{figure}

A collimated $^{133}$Ba source is mounted onto a horizontal scanning system which has two level beams that are perpendicular to each other, shown in the right panel of Figure~\ref{fig:setup}. Via the beams, the collimated source is controlled by two stepping motors. The source is normally set to scan the detector top surface along two perpendicular diameters, and it can also be modified to scan the side of the detector vertically. Detector responses, such as the rise time and pulse shape, are studied for each scan position. For example, see the left panel of Figure~\ref{fig:results}.\\

Thoriated welding rods ($^{232}$Th source) are used to study detector PSD performance. The double escape peak of 2614.5 keV photons of $^{208}$Tl is at 1592.5keV and it is of particular interest for the PSD study.  In such events, an electron-positron pair is created by pair production, and, due to very short mean free path in Ge, they deposit their kinetic energies at locations very close to each other. Since the two photons created by positron annihilation typically escape from the detector, such double escape events are single-site events, good proxies for the double beta decay signals.  At the same time, most of the full energy photon peaks with similar energy, {\it e.g.} 1588.2~keV and 1630.7~keV photons from $^{228}$Ac and 1620.5~keV photons from $^{212}$Bi, undergo multiple Compton scattering and thus are multi-site events.  The separation between SS and MS events is achieved by looking at the variable of peak amplitude to total energy or A/E, which is the ratio of the maximum current, calculated from the time differentiation of the recorded charge waveforms, to the event total energy~\cite{AoverE}. With total energy deposited at multiple stages, MS events generally have smaller maximum current for the same energy, {\it i.e.} smaller A/E. By applying appropriate A/E cuts, we are able to reject about 90\% of the MS events, while retaining about 90\% of the SS events in this energy range, as shown in the right panel of Figure~\ref{fig:results}.\\

\section{Status of Detector Acceptance and Characterization}
To date, we have received and performed acceptance and characterization tests on 28 ORTEC p-type point contact HPGe detectors at the SURF underground lab. These detectors have shown satisfactory characteristics, such as excellent energy resolution, low leakage current and good PSD performance. Data-simulation comparisons have shown dead layers that match manufacture specifications. Detailed analyses of characterization data have been established and are being improved. In short, an extensive acceptance and characterization program for detectors used in the \mj\ \dem\ is in progress.

\section{Acknowledgments}
We acknowledge support from the Office of Nuclear Physics in the DOE Office of Science, the Particle Astrophysics Program of the National Science Foundation, and the Russian Foundation for Basic Research. We acknowledge the support of the Sanford Underground Research Facility administration and staff.

\label{}



%% References


\bibliographystyle{elsarticle-num}



\end{document}